# Descriptions of women are longer than that of men: An analysis of gender portrayal prompts in Stable Diffusion.

*Yan Asadchy[1]\* & Maximilian Schich[2]*

[1]*School of Humanities, Tallinn University, Tallinn, Estonia*
[2]*Baltic Film, Media and Arts School, Tallinn University, Tallinn, Estonia*

## Abstract

Generative AI for image creation emerges as a staple in the toolkit of digital artists, visual designers, and the general public. Social media users have many tools to shape their visual representation: image editing tools, filters, face masks, face swaps, avatars, and AI-generated images. The importance of the right profile image can not be understated: It is crucial for creating the right first impression, sustains trust, and enables communication. Conventionally correct representation of individuals, groups, and collectives may help foster inclusivity, understanding, and respect in society, ensuring that diverse perspectives are acknowledged and valued. While previous research revealed the biases in large image datasets such as ImageNet and inherited biases in the AI systems trained on it, within this work, we look at the prejudices and stereotypes as they emerge from textual prompts used for generating images on Discord using the StableDiffusion model. We analyze over 1.8 million prompts depicting men and women and use statistical methods to uncover how prompts describing men and women are constructed and what words constitute the portrayals of respective genders. We show that the median male description length is systematically shorter than the median female description length, while our findings also suggest a shared practice of prompting regarding the word length distribution. The topic analysis suggests the existence of classic stereotypes in which men are described using dominant qualities such as "strong" and "rugged". In contrast, women are represented with concepts related to body and submission: "beautiful", "pretty", etc. These results highlight the importance of the original intent of the prompting and suggest that cultural practices on platforms such as Discord should be considered when designing interfaces that promote exploration and fair representation.

## Introduction:

Visual culture experienced a sustained rise with the proliferation of visual social media platforms, such as TikTok and Instagram. Over the last decade, the shifting toward new digital media platforms has altered how people create, share, and consume images, providing an interactive playground for co-creation and reuse (Manovich, 2009). Social media platforms with a strong visual focus have played an important role in this transformation. As a result, images have become even more central to personal expression, marketing (Gretzel, 2017), activism (Poell, 2014), and more. According to Bond Internet Trends, visual social media are still on the rise, with Instagram, Facebook, and WhatsApp stories having 1.5B daily active users globally and showcasing a double increase compared to the previous year (Meeker, 2019). This massive visual engagement indicates a cultural shift where people increasingly communicate and express themselves through visual content rather than written forms. The development and adoption of AI technologies, such as direct-use diffusion models (Midjourney, Stable Diffusion) and AI-enabled tools, such as Photoshop, Discord, and playground.ai, further facilitate a fertile environment for producing and consuming visual culture. The apparent quality of the models' output invited larger audiences and further pushed the adoption of generative tools (Gozalo-Brizuela, 2023). By simplifying and reducing the creative process to "prompting" — a set of short natural-language commands that work as input for the generative AI to produce an image (Oppenlaender, 2023), AI tools encourage professionals and amateurs to engage more deeply with visual culture. This intuitive and natural interaction further contributes to the mass adoption of AI tools and the proliferation of ideas. The availability, speed, and simplicity of image creation have already produced more images than 150 years of photography (Valyaeva, 2023). Meanwhile, this mass activity of visual AI prompting constitutes a natural feedback loop that effectively enables individual intuition of description quality in the user, which can translate into an *aesthetic measure* (cf. Birkhoff, 1933; Karjus et al. 2023) of *ekphrasis* that is an evaluation of Horatio famous postulate "ut pictura poesis", i.e. the equivalence of painting and poetry (Horatio, 1926).

The design of platforms such as Discord nurtures public sharing and provides quick tools to reuse and enhance images of others, creating ecological niches of visual information and aesthetics. The comparative analysis of the use of two Generative AI platforms (Stable Diffusion and Pick-a-Pic) reveals that rather than focusing on developing new prompts, users opt for the creation of image variants, resulting in silos of visual culture (Torricelli et al., 2024). Such platform affordance and user practice of reuse can not only be a limitation, but a fertile ground for establishing new and sustaining existing stereotypical portrayals, shaping modern digital practices. The adoption and mass use of such practices and affordances may open new prospects for self-expression while posing important ethical and societal questions regarding representation (Scheuerman, 2020).

Recent studies in digital media ethics emphasize how AI-generated content can unintentionally perpetuate harmful stereotypes regarding race, gender, ethnicity (Nicoletti & Bass, 2023; Gorska & Jamielniak, 2023; Jääskeläinen & Åsberg, 2024), and political affiliation (Heikkilä, 2023) or even exclude certain groups, rendering them effectively invisible. Gorska and Jamielniak (2023), for instance, found a massive overrepresentation of men (76%) compared to women (8%) in the AI-generated images of professionals in the fields of engineering, research, law, and medicine. These findings reflect inherently biased datasets used for training AI models, which skew depictions of race and gender (Sham et al., 2023). These biased representations emphasize preexisting stereotypical depictions of men and women that are present on social media platforms. Reductive portrayals of women as embodying feminine traits like nurturing, compassion, and sociability and men as inherently aggressive, detached, and self-reliant trap both genders within cultural stereotypes that are rather exaggerated than realistic. Women's bodies are valued only if they are unusually tall, thin, and flawless, while their ideal character is sexualized, weak, submissive, and childlike. Men's bodies are admired when muscular, tall, and rugged, with perfect personalities: athletic, dominant, aggressive, sexual, and powerful (Newton & Williams, 2011). These stereotypes are particularly important within the context of visual representation since visual information is the most effortlessly remembered and propagated (Graber, 1990), with stereotypical depictions being recognized most efficiently (Palmer, 1999).

Our goal in this research is to closely examine the intent in generating images of men and women through the lens of text used in prompts, i.e. the user-generated input text that sparks the machine generation of images. We are particularly interested in how different the language and structure of prompt descriptions of images of men and women (RQ1). What words define the descriptions of men and women in our chosen DiffusionDB dataset (RQ2)? What are the stereotypical portrayals of men and women (RQ3)? We will answer these questions using text analysis and statistical methods described in greater detail in the methods section below.

Our findings suggest a predisposition and demand for AI to produce stereotypical depictions of men and women. We hope our findings will become grounds for further discussions on AI-generated visual culture, balanced/representative datasets, gender representation, the formation of stereotypes, and the role image-generating software interfaces play in it.

# Data:

DiffusionDB[1] is the pioneering large-scale dataset documenting text-to-image generation, encompassing 14 million images synthesized by Stable Diffusion (Rombach et.al, 2022) produced with the help of users of the Discord social media platform (Discord, 2024). This dataset represents a significant advancement in the documentation of generative model output, capturing the dynamic interaction between text prompts and image outputs based on considerable amounts of real-world user input. Hosted on the Hugging Face repository, DiffusionDB enables research into various domains, including prompt-model interactions, deepfake detection, and the design of human-AI interfaces.

DiffusionDB is available in two distinct subsets tailored to different research needs: a smaller dataset, DiffusionDB 2M, which contains 2 million images and 1.5 million unique prompts, and DiffusionDB Large, which encompasses 14 million images and 1.8 million unique prompts. Images in DiffusionDB 2M are stored in PNG format, while DiffusionDB Large utilizes a lossless WebP format. Each image file in the dataset is uniquely identified by a UUID (Version 4), ensuring a distinct reference across the dataset. JSON files within directories contain key-value pairs linking images to their respective prompts and generation parameters. This study uses all 1.8 million prompts from the DiffusionDB Large Database.

## Metadata

The DiffusionDB Large metadata table provides comprehensive details about each image, such as the filename of the image, the textual prompt used for image generation, dimensions of the image, an anonymized user identifier, the timestamp of image generation, a likelihood score of the image being Not Safe For Work (NSFW) as predicted by the LAION NSFW detector (Schuhmann et al., 2022), a likelihood score of the prompt being NSFW, as well as other technical metadata that we did not use in this research.

---

[1] Stable Diffusion: https://huggingface.co/datasets/poloclub/diffusiondb

DiffusionDB is publicly available, and Stable Diffusion's open-source nature and the generated images' CC0 1.0 Universal Public Domain Dedication facilitate broad and unrestricted use. The dataset excludes explicit personal information, anonymizing user identities to prevent potential misuse.

While DiffusionDB captures a broad spectrum of image styles and prompts, the dataset originates from a specific community of early adopters, which may have introduced specific biases. The prompting style prevalent within this dataset might only partially represent novice users or specific domains requiring specialized knowledge. Additionally, the individuals who generated images will likely be young and tech-savvy internet users, which is in line with the Discord platform user demographic. As such, the findings presented here cannot necessarily be extrapolated to the behavior of the general population, even though the scale of this *iconographic* analysis (Panofsky, 1939; Schich, 2019) may perhaps seem unprecedented or at least considerably beyond the scale of conventional literature in art history or ethics in visual communication (Lester, 2018).

# Methods:

## Data preparation

We use RegEx to identify prompts that depict either men or women by first identifying necessary keywords to filter the prompts ("male", "man", "boy", and "guy" for men, and "female", "woman", "girl", and "gal" for women). We further filter the dataset and label images and prompts with the respective apparent gender. To isolate images containing groups of people, we remove prompts that contain a combination of any of these words from our sample. We also use Regex to count the number of words in each prompt. The resulting set of prompts undergoes a series of traditional text analysis transformations. First, we transform the text to lowercase, strip the leading and trailing spaces, and then delete any multiple spaces. Lastly, we removed short words (less than two symbols). We used the Natural Language Processing Toolkit NLTK (Bird, 2006) to remove irrelevant stop-words and punctuations to prepare the data for the training and testing.

## TF-IDF

We process the resulting collection of prompts using the term-frequency inverse document frequency (TF-IDF) method (Salton et al. 1973), assigning a weight to each word in each prompt. TF-IDF is a reliable method for identifying important words and terms across multiple documents (Ramos, 2003). In this study, we use TF-IDF on single words or unigrams. To test the accuracy of labeling, we perform a simple linear regression analysis.

## Close reading and manual clustering

Identifying the most frequent terms describing men and women in the prompts is insufficient to derive meaningful conclusions about their meaning and role in constructed representations. Some terms can have different meanings when situated in a range of contexts, such as computer graphics, social media, digital art, and especially games, considering that images were created on a platform initially oriented at gamers. To address this, we further analyze and categorize the most important terms by closely reading a selection of prompts to understand the context of their use.

Finally, we perform an expert annotation to identify dominant themes among the frequent words used in prompts. We chose a manual approach in favor of automated approaches such as *latent Dirichlet allocation* (Biel et al., 2003) or *Latent semantic analysis* (Dumais, 2004) for multiple reasons. First, the resulting set of words is evidently simplistic, and using quantitative methods won't bring significant improvements. Second, the meaning of some words is contextually dependent, and quantitative methods might overlook these nuances. Third, the authors are experts in digital culture, art history, and AI tools for image generation. Considering these points, we opted for quantifying summary statistics, complemented by qualitative analysis using manual sorting and systematic hermeneutic observation, to derive meaningful categories and interpretations of results.

# Results

## Prompt properties

First, we find a substantial overrepresentation of women in our sample, with 32.2% of prompts describing eventually male versus 67.8% eventually female depictions, as visible in Figure 1a. Since the dataset does not contain any information regarding the demographics of the Discord users, we unfortunately cannot answer some obviously related questions. Specifically, how does this overrepresentation relate to the user's gender, or is there a difference between how men and women construct prompts to create images of men and women? However, we can look at the general trends of their depiction. Our analysis of word prompt lengths shows patterns in how users produce gendered images using varying prompt lengths (Figure 1b). The maximum prompt length is 480 words, the minimum length is 0, the median length is 23, and the standard deviation is 16 ($Max_{length}$ = 480, $Min_{length}$ = 0, $Median_{length}$ = 23.0, StD = 16).

**Fig 1. The description length of women is systematically longer than men's with the tailed length distribution and a meaningful median. Very few Discord users are accountable for the longest prompts.**

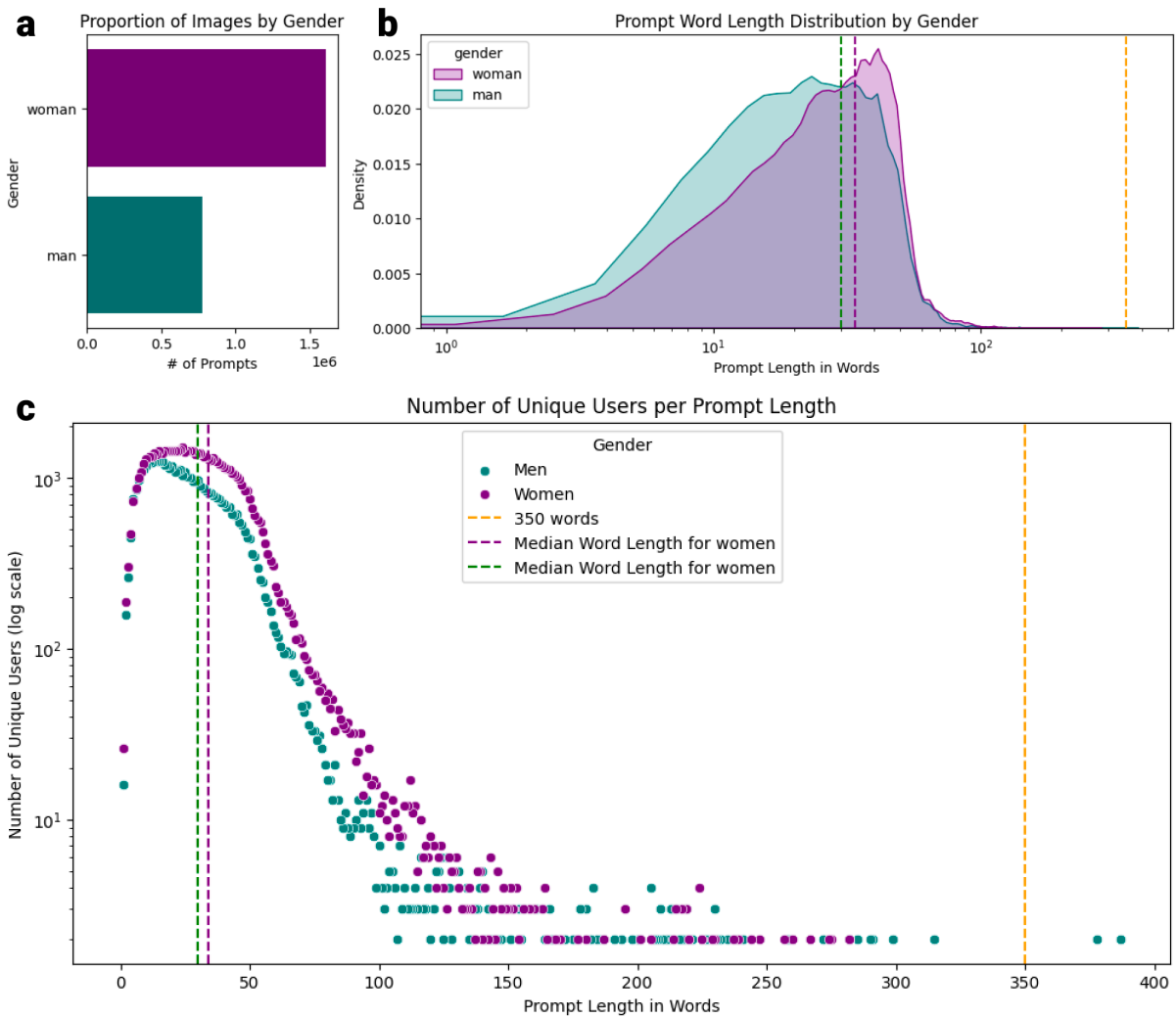

*A.* Images of women are overrepresented compared to men (1,606,013 vs 779,200).
*B.* Images of men are described in fewer words than those of women. As the length of the prompt increases, the distributions exhibit a sharp decline of around 50-80 words in length, highlighting the end of the optimal prompt length. The orange dashed line indicates the prompt length limit of 2000 characters (approx. 350 words) specified by the Discord platform.
*C.* Distribution of unique users by prompt length for images depicting men and women. The x-axis indicates prompt length in words, while the y-axis shows the number of unique users on a logarithmic scale. The trend shows that the highest number of unique users prefer prompts around 20-50 words to describe images of men and women, with a significant decline in unique users as prompt length increases. Despite differences in the median prompt length, the overall prompt construction pattern is similar across gendered images, suggesting a shared preference for concise prompts.

This distribution shows an inverse relationship between prompt length and the number of unique users who produced a prompt of a specific length. We use the number of unique users to avoid a situation where a small number of users who generate a large number of images with a certain prompt length skews the distribution. It is seen from the figure that shorter prompts are consistently used to describe men across the whole distribution. Discord has a limit of 2000 characters (approx. 350 words) per message, which limits the amount of details and contextual information users can put in the prompt. However, we did not find systematic evidence of users hitting the hard character limit and using shorter words to construct the prompt, with only some exceptions in the dataset ($Max_{char}$ = 1923, $Median_{char}$ = 152, $std_{char}$ = 114).

The number of unique users peaks for relatively short prompts (around 20-50 words) and then rapidly declines as prompts get longer, suggesting that only a few users are willing to experiment with longer prompts. While the overall pattern is similar for images featuring either gender, pictures of women generally were made by a higher number of unique users across most prompt lengths, particularly noticeable in the peak range and for moderately long prompts (50-150 words). This finding should perhaps be expected considering the ratio between the number of images depicting women and men in our sample ($N_{woman}$ = 1,606,013, $N_{man}$ = 779,200).

Compared to the messages on other text mediums like YouTube comments (Sobkowicz et al., 2013) and Twitter posts (Gligorić et al., 2018), the distribution of prompt length differs in its characteristics. In the case of YouTube, where comments have a 10,000-character limit, there is plenty of space for expressing one's opinion or ideas, as reflected by a bell curve distribution of message length. Prompt length distribution on Discord, however, is more skewed. Meanwhile, we also find a meaningful median or non-zero peak in our frequency distribution, highlighting the existence of an optimal prompt length needed to produce a desired image with StableDiffusion. Interestingly, the users of Discord do not participate in maximizing the prompt length to reach the limits imposed by the platform. Neither do we find evidence for the existence of two distinguished ways of prompting, like on Twitter, where the bell curve has a secondary spike towards the character limit as the result of cramming, essentially optimizing long messages to fit into one post. This finding suggests that while the prompt character length limit is strict, our Discord users do not optimize their prompt writing practice to fit as much contextual information as possible. This is particularly interesting, considering that models such as Stable Diffusion work better when given more detailed descriptions (Betker et al., 2023).

## Word analysis

Since there are significantly more prompts associated with depictions of *women* than with those of *men*, this class imbalance could lead to a biased performance of our analysis when using a model classifier. To address this imbalance, we employed an undersampling technique, reducing the number of samples in the *woman* class to match the number of samples in the *man* class. This way, we can be sure the model will not disproportionately favor the majority class based on a priori class imbalance.

After undersampling, we retrained a logistic regression classifier. We evaluated the regression model using accuracy, precision, recall, and F1-score. We observed balanced performance across both genders: precision (men = 0.76, women = 0.79), recall (men = 0.80, women = 0.75), f1-score (men = 0.78, women = 0.77), support (men = 0.21474, women = 0.21423) (update the numbers later).

Additionally, we visualized the most important words (n = 25) for each gender by plotting the weights assigned by TF-IDF to each word, which provided insight into the specific language patterns most indicative of male and female prompts used by the Discord users of the Stable Diffusion. A more in-depth analysis of words is provided in the follow-up section. Still, already from Figure 2, it can be seen that words such as "masculine" and "bearded" were more predictive of male prompts, while terms like "beautiful" and "sexy" were strongly associated with female prompts. Further below, we closely inspect words associated with men and women and the results of expert manual clustering.

**Fig. 2. Stable diffusion users generated beautiful women and brutal men.**

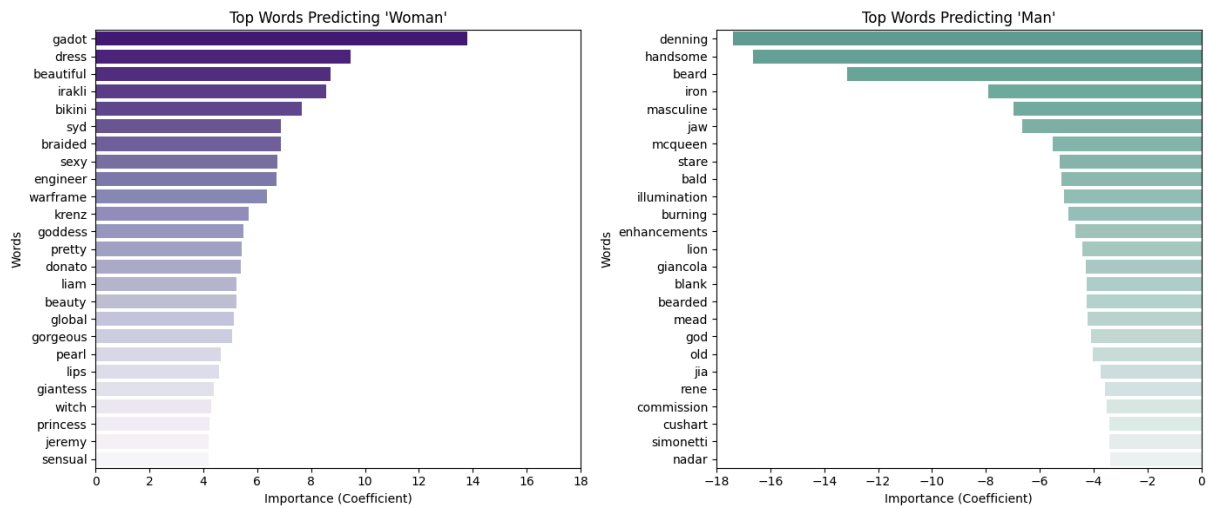

*This plot compares the top 25 most predictive words for images depicting men and women. On the y-axis, we see the coefficients assigned to each term by the TF-IDF algorithm. For better visualization, we split the coefficients into positive for women and negative for men. On the x-axis, the terms are sorted in descending order from most important to least. Prompts for images featuring men and women exhibit a similar pattern, with men's coefficients having a sharper decline.*

## Topic modelling

In this chapter, we provide the results of the manual topic modeling we performed on the words that are the highest predictors of images of men and women in our Discord-StableDiffusion dataset. We opted for manual topic modeling for several reasons, such as the small size of the top salient words selected from the TF-IDF analysis and the specifics of the terminology used in the respective prompts. The results of our manual topic modeling suggest the existence of strong stereotypical topics that are aligned with previous research on gendered representations (Newton & Williams, 2011; Palmer, 1999). Below, we provide a list of these topics or themes, with examples of words that constitute them, based on prompt examples from the DiffussionDB dataset.

### Appearance & Physical Attributes:

In our dataset, the physical appearance of women is described with the following words: *'beautiful', 'bikini', 'braided', 'sexy', 'pretty', 'beauty', 'gorgeous', 'pearl', 'lips'* and *'warframe'*, while men's physical appearance is described with stereotypical traits of masculinity (Mosse,

1998): *'handsome', 'beard', 'jaw', 'bald', 'old'* and *'stare'*, which is in line with the previous research that conventional images of men and women have different prominent characteristics. Where women are portrayed as supposedly 'sexy', 'submissive', and 'dependent', men usually supposedly possess dominant and active traits (Williams & Best, 1990). Additionally to this list of words, the word *warframe* stands out. "Warframe" is a game famous for its distinguished visual style that blends sci-fi, cyberpunk, and space fantasy[2]. The game's organic and futuristic aesthetics create a unique biomechanic look. The female *warframes* in the game are frequently designed to emphasize highly sexualized body shapes, showcasing sleek hourglass figures and tight armor designs. Interestingly, the *warframes* are not humans but biomechanical exo-suits worn by mythically ancient warriors of unknown origin. However, despite this abstraction, they are easily mistaken for humans; thus, it doesn't distance them from more traditional and sexualized depictions of women in other games, which is why we keep them in this emerging category.

## Strengths & Powers

It is not surprising that stereotypical descriptions of men are richer in words related to strength and power attributes. Words such as 'iron', 'masculine', 'lion', 'burning', and 'god' dominate the prompts used to generate images of men. For images depicting women, the words *'goddess'* and *'giantess'* emerge as definitive and are used to describe a powerful individual, contrasting with the stereotypical image of a sensitive and submissive woman.

## Roles & Professions

These terms indicate men's and women's involvement in certain professions or activities, such as 'engineer', 'witch', and 'princess' for women. This finding is particularly interesting in light of the previous work of Gorska and Jemielniak (2023), who found the underrepresentation of engineers among images of women. Notably, there is no mention of men's professions, which might be a case of focus on faceism—a practice of focusing on a man's face without providing much context regarding the occupation or status of the depicted person (Archer et al., 1983).

## Artists style

Specifying the artist's name in the prompt is a common practice in prompt engineering that drives the outcoming picture closer to the desired outcome. A subject of copyright infringement, companies like Stable Diffusion used many copyrighted materials to train their models, allowing creators to parody their beloved artists' styles. Injecting the prompt with the artist enables the artist to infuse the outcome image with colors, brushstrokes, prominent anatomical and facial features, outfits, and accessory designs. In our study, we identified a range of influential artists whose styles were excessively used to define the look of men and women generated by the users of Stable Diffusion. In particular, the images of men are associated with the artistic styles of *Guy Denning*, who is famous for his portraits featuring long and contrasting brushstrokes, as well as *Donato Giancola, Ryan Jia, Irakli Nadar, Krenz Cushart, Rene Magritte, and Syd Mead*, a visual futurist famous for his epic sci-fi landscapes. Similarly, women are depicted using the styles of *Irakli Nadar*, a Georgian portrait artist famous for soft and gentle anime-style portraits of women, and *Liam Sharp* - a

---

[2] https://www.warframe.com/game

British comic book artist famous for hyper-detailed images of superheroes and Wonder Woman in particular, *Syd Mead*, *Donato Giancola*, *Jeremy Mann*, and *Krenz Cushart*. It is worthwhile noting that our approach of applying TF-IDF, followed by manual inspection of terms, lends itself beyond the scope of this paper, to identify potential bulk copyright violations by machine learning companies that profit from AI image generation.

## Celebrities

Actors and models have always been beauty standards and ideals one could strive to become. Among the most influential words in prompts, we found references to celebrities, which users of StableDiffusion chose as the ideals of masculinity and femininity. Among the top 25 prompt strings for images of women is *Gal Gadot*, an actress known for her portrayal of Wonder Woman, and for men, it is *Steve McQueen*, a well-known actor.

## Technical

Good prompting results in predictable images. To do so, prompt engineers use a wide range of technical terms to increase the quality of the output image. *Global illumination* is one such term used in computer graphics to describe a set of techniques that more realistically simulate how light interacts with surfaces. Pictures generated with this word in the prompt have indirect lighting caused by reflections, refractions, and light scattering across multiple surfaces. This technique and other 'contrast enhancements' make images more natural and colorful. *Commission* is another technical term used in prompting along with the artists' names. While it does not affect the output value, it is believed to make using the artist's name in the prompt more "ethical." This practice has no legal basis, and Adobe's content policy, for example, explicitly forbids artists' names from being used in prompting.[3]

# Conclusions

In this study, we used statistical methods to analyze and classify text prompts to understand what images of men and women were generated using the StableDiffusion model. We specifically focus on the underlying language that differentiates between prompts associated with men and women. We used a StableDiffusion dataset of images and prompts, labeling each pair based on what they depicted: a man or a woman. We then employed the natural language processing (NLP) method to vectorize and model the language features that predict gender. Using logistic regression, we trained a classifier to predict whether a given prompt was associated with the labels "man" or "woman" we gave them earlier. We analyzed the coefficients of each word in the regression model to determine which terms were the strongest predictors for each gender.

We looked at the 1,812,994 prompts written by 10,380 unique users to generate images of men and women. Our findings suggest that most users prefer to write relatively short prompts, around 20-50 words long. Very few users write longer prompts exceeding 150 words. We also found that male depictions are generally described by shorter prompts and a larger number of unique users generated images of women, which was expected considering the overrepresentation of female depictions in the dataset. Despite variations, the overall pattern of the unique user frequency distribution over prompt length appears similar for both genders, suggesting a shared preference for prompt lengths.

---

[3]https://helpx.adobe.com/stock/contributor/help/updated-artist-name-guidelines.html

By looking closely at the words used in the analyzed prompts, we found that Discord users of the Stable Diffusion service generated rather stereotypical depictions of men and women. Using TF-IDF, we found the most frequent words used within prompts. Among these words, we see the dominance of traditional masculine traits revolving around strength and brutality, while women are described using words focused on beauty and cuteness.

Previous research on other text-to-image models (Midjourney) highlighted a similar reinforcement of harmful beauty norms, where "beauty" is often represented by young, white, heteronormative women (Jääskeläinen & Åsberg, 2024). In our study, we find a similar trend and highlight that users of Discord requested StableDiffusion to produce images of evidently stereotypical portrayals of men and women.

In conclusion, this study underscores the significance of understanding how users use generative AI tools, particularly on platforms like Discord, and how they shape gender portrayals through textual prompts. Our analysis of over 1.8 million prompts reveals ingrained cultural stereotypes, with men typically described in dominant, action-oriented terms, while women are more often associated with physical appearance and submission. These findings raise essential questions on how cultural practices and biases are solidified on these platforms by generative AI and how they may unintentionally reinforce traditional stereotypes.

From the perspective of intercultural communication, this study highlights the need for AI systems and interfaces to be designed with respect to cultural and social practices. The biases uncovered here illustrate how cultural representations can be reinforced when AI systems rely on user-generated content, particularly in global environments such as Discord.

Finally, the research offers practical implications for designing fair and inclusive AI systems. Our work suggests that having balanced datasets that make representative AI models is not enough. It is necessary for designers of generative AI tools to pay greater attention to the prompting mechanisms and interface designs that shape user behavior to avoid the production of overly stereotypical depictions. By raising awareness, we hope that our work can help prioritize fairness and inclusivity by designing generative AI tools that positively impact intercultural communication while battling the perpetuation and reinforcement of harmful gender stereotypes.

# Limitations

While this analysis provides valuable insights into the production of culture, it is important to note its limitations. The data do not account for the gender of the content creators, who are the users of the Discord platform, nor do they consider the specific themes or contexts of the prompts beyond gender categorization. Future research should explore these additional variables to provide a more complete understanding of gender stereotypes in AI-generated images.

Focus on only one platform brings its limitations. While DiffusionDB is one of the largest repositories of AI-generated images, it is essential to account for other services and interfaces that allow users to create, store, and reuse the visual content before we can make generalized claims about the emerging culture of AI image production.

Furthermore, a deeper qualitative and quantitative analysis of the prompts' content on various platforms could offer new insights into cultural differences and how the interface can affect user preferences and behavior. We hope our paper provides inspiration in these directions.